# An Investigation of Thermal Properties of Cu-Au Janus Nanoparticles

M. A. Cebeci[1], H. Zor Oğuz[2], S.Ozdemır Kart[2], H. H. Kart[2]
[1] Galatasaray High School, Galatasaray University, İstanbul 34430, Türkiye.
[2]Department of Physics, Pamukkale University, Kınıklı Campus, 20017, Denizli, Türkiye.

**Abstract**

In this paper, the thermal and structural properties of Cu-Au (Copper-Gold) Janus nanoparticles with a diameter of 5 nm are investigated by using molecular dynamics (MD) simulations within the interactions defined by the many-body embedded atom model (EAM). A set of nanoparticle models has been constructed, with varying Cu and Au ratios. MD method is carried out to calculate the melting temperature, heat capacity, radial distribution function (RDF), Lindemann index, mean square displacement (MSD), and diffusion coefficients of these models. The findings demonstrate that nanoparticles rich in Cu exhibit a higher melting temperature and more defined phase transitions. In contrast, structures rich in gold exhibited reduced melting temperatures and showed surface-initiated melting behaviours. MD study highlights that the thermal stability and atomic mobility of Cu-Au Janus nanoparticles depend on the composition ratio and the dispersion of materials.

1. **Introduction**

Nanotechnology has advanced dramatically in recent years and made it easy to explore and customize the physical, chemical, and biological properties of nanomaterials with higher precision. Nanoparticles (NPs) have been at the centre of these advances, contributing to diverse fields such as medicine, environmental science, energy systems, electronics, and materials research (Khan, 2019, Malik, 2023, Srinivasan, 2016, Zhong, 2025). The specific behaviour exhibited by these systems is predominantly attributable to their elevated surface-to-volume ratio, size-dependent structural properties, and quantum confinement effects that become manifest at the nanoscale (Blackman, 2018).

Janus nanoparticles (JNP) are of particular interest in this regard. These nanoparticles are named after the Roman god Janus, who is often depicted with two faces looking in opposite directions.  The notion of JNP was initially proposed in 1991 by Pierre-Gilles de Gennes during his Nobel Prize address (de Gennes, 1992).  Since then, JNPs have become a prominent example of next-generation nanomaterials science. The most outstanding feature of JNPs is their capacity to incorporate two distinct materials in different regions of their surface. This characteristic allows JNPs to exhibit both hydrophilic and hydrophobic, cationic and anionic, or magnetic and fluorescent properties (Li, 2022, Kaewsaneha, 2014, Li, 2016, Niedner, 2024). This multifunctional character makes JNPs highly suitable and strategic materials not only in biomedical applications but also in many advanced technology areas such as emulsion stabilization, targeted drug delivery, imaging technologies and photocatalytic reactions (Vihal, 2025).

Biomedical applications of JNPs are also crucial. The chemical bifaciality makes these particles applicable both as carriers of therapeutic agents and as imaging agents. For instance, JNPs with magnetic-fluorescent or hydrophilic-hydrophobic structures can be used simultaneously for magnetic resonance imaging (MRI) and targeted drug delivery. Such particles offer a significant advantage in multimodal approaches such as photothermal therapy (PTT), photodynamic therapy (PDT), chemotherapy and magnetocaloric therapy (Zhang, 2021).

The utilisation of JNPs is not only limited to medical applications but also includes catalysis, environmental technologies, micromotor systems and energy conversion applications (Gao, 2014, Wang, 2021, Chauhan, 2018). The presence of asymmetric surfaces on the material under investigation has been demonstrated to result in the achievement of selective catalytic activity at a range of surface sites (Jia, 2016). The employment of JNPs in the domains of steerable microsystems design and multi-step catalytic reactions within the same structure is facilitated by the chemical differences on their surfaces (Vafaeezadeh, 2022).

The behaviour of JNPs at liquid-liquid interfaces enables them to act as stabilisers, particularly in emulsion systems (Glaser, 2006). It is demonstrated that these structures facilitate the formation of more stable Pickering emulsions by reducing surface tension. In the domains of cosmetics, food, petrochemicals, environmental engineering and pharmaceuticals, this property finds application in the control of emulsion-based systems(Zhang,

2022, Chen, 2020). The versatile surface design of JNPs plays a significant role in several applications such as targeted substance recovery, selective separation and controlled molecule release (Su, 2019, Rahiminezhad, 2020, Dehghani, 2018).

Janus materials are also used in the structures of advanced materials (Yang, 2017; Zhang, 2021; Pradhan, 2007). In the field of materials science, the multi-material nature of JNPs has rendered these innovative nanomaterials a viable solution not only for traditional applications but also for advanced technology applications, ranging from smart surfaces to programmable systems and shape-memory materials (Zhang, 2016, Shah, 2015, Yang, 2017, Safaie, 2020). Through their interactions with supramolecular structures, JNPs pave the way for the construction of hierarchical multistage assembly mechanisms and simultaneously enable controllable self-assembly processes that can be directed by external stimuli. This facilitates the integration of advanced material designs that combine high structural flexibility with functional responsiveness within a single system (Kang, 2018, Reguera, 2020, Li, 2021, Zhang, 2018). For instance, JNPs, which are biocompatible and drug-loaded on one side and magnetic on the other, are used as multifunctional nanoplatforms in the biomedical field, offering both targeted drug delivery and magnetic guidance capabilities (Xing, 2018, Feng, 2019, Shao, 2016, Ali, 2024).

In recent years, there has been a significant increase in the number of experimental studies demonstrating the potential applications of copper (Cu)-gold (Au) JNPs. These applications show that Janus nanostructures with Cu–Au compositions can be used in advanced treatment methods because they exhibit photothermal, antibacterial, and therapeutic functions. Indeed, Au–Cu JNSs, which combine photothermal plasmonic activity and the antibacterial properties of Cu in a Janus architecture, have been found to have high potential for broad-spectrum antibacterial treatment and biomedical applications (Yang, 2024). Furthermore, controlling interfacial energy through strong thiol ligands enables the conversion of Au–Cu core–shell structures into asymmetric Janus structures, offering new opportunities for precisely designing bimetallic nanostructures (Fan, 2022). In short, studies on Janus nanoparticles reveal classifications consisting of polymeric, inorganic, and hybrid structures, as well as various production strategies and broad biomedical application potential. These studies demonstrate advances in the field by showing how Janus nanoparticles integrate different functional properties, arising from their asymmetric structures, into a single system (Zhang, 2021).

These experimental findings have also been supported by molecular dynamics (MD) simulations, which help us understand the atomic-level behavior of Cu–Au JNPs. MD simulations reveal the thermal stability, alloying tendencies, surface segregation, and atomic rearrangement processes of Cu–Au Janus nanoparticles. These simulations provide results that are consistent with experimental observations. For example, Chepkasov et al. (2018) have reported high structural stability in Janus-type structures within specific temperature ranges. They compared the alloy, core–shell, and Janus configurations of 5-nm Cu–Au nanoparticles to reach this conclusion (Chepkasov, 2018). Similarly, another study used molecular dynamics simulations to investigate the effect of atomic ratio and interface structure on the surface evolution of Janus Cu–Ag nanoparticles during sintering. The study showed that different atomic arrangements and interfaces affect the unique sintering behavior of Janus structures and their transition to Cu/Ag alloys when heated (Liang, 2020). Studies of the molecular dynamics of Fe–Cu bimetallic nanoparticles have shown that phase transitions and structural order depend on the morphology (e.g., random alloy, core–shell, or Janus) and cooling rate (Kumar, 2019). Similarly, molecular dynamics simulations of Ni–Au Janus nanoparticles show that the Au/Ni ratio and particle size influence atomic segregation, coalescence processes, and the final morphology (e.g., dumbbell, Janus, or eccentric core–shell structures) when heated (Li, 2021). Additionally, molecular dynamics simulations of the formation of Fe–Cu bimetallic nanoparticles via high-speed collisions reveal that the formation of phase-segregated Janus nanoparticles is determined by different size, temperature, and velocity conditions. These simulations show good agreement with experimental results (Tsukanov, 2019).

## 2. Computational Method

In this study, Janus Cu-Au nanoparticles (NPs) with a diameter of 5 nm have been modelled to maintain a face-centred cubic (fcc) crystal structure. The NPs have been placed in a 200×200×200 Å simulation box for molecular dynamics (MD) simulations. The construction of the Janus configurations involved the division of the spherical nanoparticles into two hemispheres, creating asymmetric compositions through the strategic placement of Cu and Au atoms in discrete regions. The model systems have been designed as five different structures in which the ratio of Cu to Au atoms has been adjusted by decreasing the number of Cu atoms and increasing the number of Au atoms. As shown in Figure 1, these structures were modelled with ratios of $Cu_5:Au_1$, $Cu_4:Au_2$, $Cu_3:Au_3$, $Cu_2:Au_4$ and $Cu_1:Au_5$. Figure 1 shows the visualisations of the Janus structures rendered using the Open Visualisation Tool (OVITO) (Stukowski, 2010).

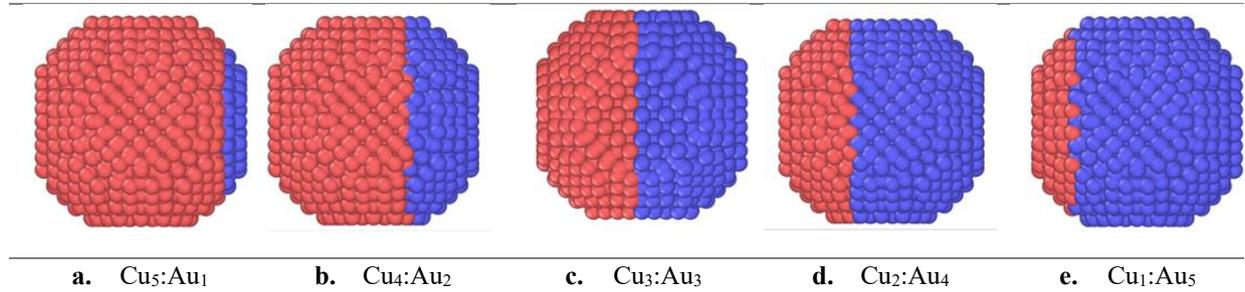

a. $Cu_5:Au_1$    b. $Cu_4:Au_2$    c. $Cu_3:Au_3$    d. $Cu_2:Au_4$    e. $Cu_1:Au_5$

**Figure 1.** 5 nm simulation images of model systems (where magenta colour represents Cu atoms and blue colour represents Au atoms).

Our model systems consist of a total of 3589 atoms. The number of atoms for each material in these models shown in Table 1. Molecular dynamics (MD) simulations have been performed to understand the details of the melting mechanism of the model systems. These simulations have been performed using the LAMMPS software package. Embedded Atom Method (EAM) type potentials, specifically developed for Cu and Au atoms (Daw, 1984; Zhou, 2003), have been used as input parameters to the program to describe the interatomic interactions.

**Table 1.** Comparison of the number of nanoparticles (NPs) and melting temperatures of 5 nm model systems with experimental data.

| Model Systems (Cu:Au) | # of atoms Cu | Au | $T_m$ (K) Simulation | $T_m$ (K) Experimental |
|---|---|---|---|---|
| $Au_6$ | 0 | 3589 | 1060±10 K | |
| $Cu_1:Au_5$ | 174 | 3415 | 1000±10 K | |
| $Cu_2:Au_4$ | 1105 | 2484 | 1010±10 K | 1358 K (Cu) |
| $Cu_3:Au_3$ | 1673 | 1916 | 1080±10 K | 1338 K (Au) |
| $Cu_4:Au_2$ | 2589 | 1000 | 1040±10 K | |
| $Cu_5:Au_1$ | 3415 | 174 | 1050±10 K | |
| $Cu_6$ | 3589 | 0 | 1100±10 K | |

According to the Embedded Atom Method (EAM) potential, the total energy of an atom in a crystalline structure consists of two main terms, as shown in Equation (1):

$$E_T = \frac{1}{2}\sum_{i \neq j}^{N} \emptyset(r_{ij}) + F_i(\overline{\rho_i}) \qquad (1)$$

Here, $F(\rho_i)$ represents the embedding energy for atom $i$, where $\rho_i$ is the electron density surrounding the atom. $\emptyset(r_{ij})$, is the pairwise interaction potential, defined as a function of the distance $(r_{ij})$ between two atoms. The EAM is highly effective in modeling the elastic properties, surface energies, cohesive energies, and other thermal properties of metals and alloy systems. It is widely employed in molecular dynamics (MD) simulations (Daw, 1993).

The integration of the equations of motion has been performed using the velocity-Verlet algorithm. A time step of 1 fs has been chosen for each molecular dynamics (MD) simulation step. For each target temperature, 500,000 MD heating steps and 500,000 MD equilibration steps have been performed. During the heating phase, a nose-hoover thermostat has been used to maintain a constant temperature. Throughout the simulations, atomic trajectories have been recorded every 50 MD steps and these data have been used to analyse the physical properties of the system. At each target temperature and every 50 MD steps, the temperature, energy and pressure values of the system have been determined. These physical properties have been obtained by calculating the average values over the last 500,000 MD steps during which the system has reached equilibrium.

The initial configurations of the systems have been optimised to achieve minimum energy using the conjugate gradient algorithm. In order to investigate the variation of thermal properties with temperature, the systems have been heated in the canonical NVT ensemble from 1 K to 2000 K in increments of 100 K. In order to determine the melting temperature with greater precision, the temperature interval has been narrowed to 10 K around the melting point.

In this study, the thermal properties of JNPs, including their melting point, heat capacity, radial distribution function (RDF), Lindemann index, mean square displacement (MSD), and diffusion coefficients based on MSD, have been evaluated as a function of temperature.

## 3. Results And Discussion

In this study, the structural and thermodynamic properties of 5 nm JNPs with different configurations have been investigated using the molecular dynamics (MD) simulation method with the many-body EAM potential. The model systems, number of atoms and the distribution percentages of atoms in different regions of the Janus structures have been presented in Table 1. The thermal properties have been analysed for JNP model systems given in Table 1.

### 3.1. Melting Point

The melting point of a material is defined as the temperature at which it transitions from the solid phase to the liquid phase, thereby establishing an equilibrium between the liquid and solid phases of the material (Shvartsburg, 2000). Consequently, the melting point is considered one of the material's characteristic properties, utilised for its description. The determination of the melting point entails the calculation of the total energy of the system at different temperatures and its subsequent analysis as a function of temperature. The region in which the total energy deviates from linearity and exhibits a sudden increase is identified as the onset of melting. These calculations are performed separately for each model system, and the melting regions are determined from the total energy-temperature graphs.

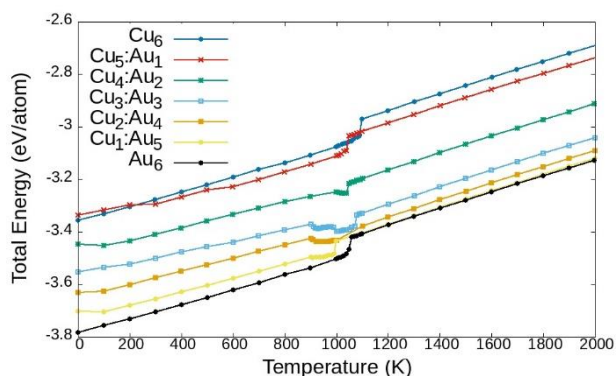

**Figure 2.** Total energy as a function of the temperature for 5 nm janus nanoparticles of various Cu:Au ratios.

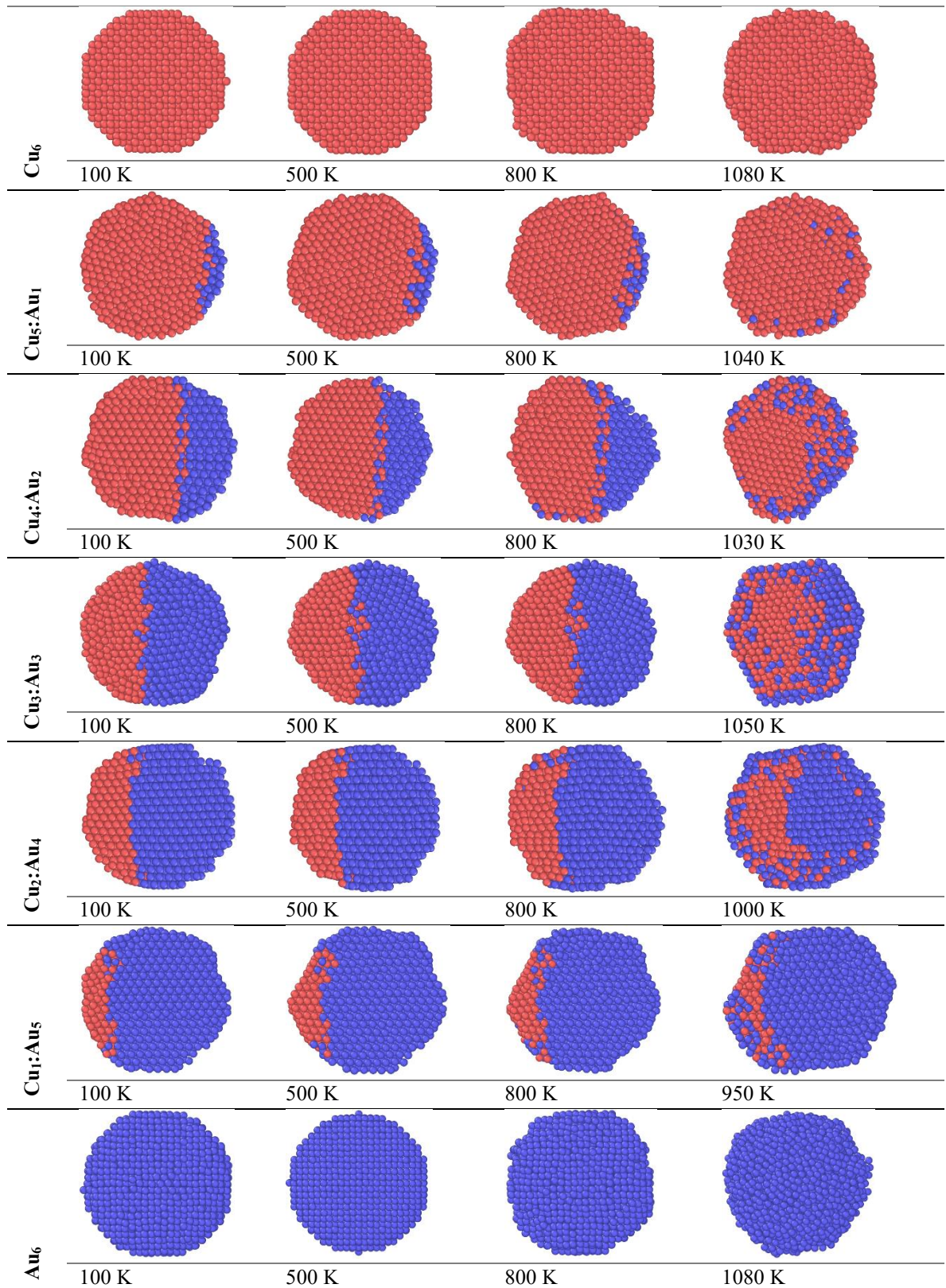

**Figure 3.** The cross-sections of the evolution of model systems during the heating process at various temperatures prior to the melting point (Magenta colour represents Cu atoms and blue colour represents Au atoms).

The total energy-temperature figures are presented in Figure 2. As shown in Figure 2, the total energy of the systems increases linearly with temperature up to a certain point, beyond which it exhibits a sudden rise. Up to this critical temperature, the nanoparticles (NPs) remain in the solid phase. The abrupt change in energy corresponds to a first-order phase transition from the solid to the liquid phase, which is defined as the melting temperature. Following this transition, as the systems are further heated, the total energy resumes a linear trend, characteristic of the liquid phase. Figure 2 shows that the total energy of the model systems increases linearly with temperature at low temperatures, which is an important result for understanding the behaviour of these systems. This indicates that the systems are in the solid phase. A sudden break and slope change is observed in the curves at a certain temperature. This point is the melting temperature. In our model systems, we have observed that the melting temperature decreases as the Cu ratios decrease. Among Janus nanoparticles (JNP), the $Cu_3:Au_3$ composition has the highest melting temperature. It is about 1080 ± 10 K. Other JNP model systems show melting below this value. In particular, the $Cu_1:Au_5$ and $Cu_2:Au_4$ models exhibit the lowest melting temperatures. Their values are around 1000 K. After the melting point, the increase in total energy with temperature is once again linear, which indicates that the system displays a stable energy increase in the liquid phase.

In order for the melting behavior of Janus nanoparticles to be better understood, a comparison has been made with pure systems of the same size. The pure Cu ($Cu_6$) system with a diameter of 5 nm has the highest melting temperature of about 1100 K, while the pure Au ($Au_6$) system has the lowest melting temperature of about 1060 K (Zor Oğuz, 2025). These results show that the thermal stability tends to increase as the Cu content in Janus nanoparticles increases and the stronger bond structures of Cu atoms are decisive in the melting behaviour. In contrast, Au-rich systems have been characterised by lower melting temperatures. This difference indicates that Janus nanoparticles containing two different metals exhibit markedly different melting behavior compared to pure Cu or Au systems. In the literature, it has been reported that the deviations observed in such bimetallic nanoparticles are due to factors such as composition-dependent surface segregation, core-shell or phase separation effects, heterogeneity in bond strengths and atomic-level disorder (Baletto, 2005, Yin, 2005, Ferrando, 2008). As seen in Figure 3, the behavior of Au towards the surface and the positioning of Cu in the inner regions lead to changes in surface energy, which significantly effects melting behavior (Wang, 2007). Therefore, the melting temperatures of Janus nanoparticles differ compared to pure phases, presenting a more complex and composition-sensitive thermal property profile.

These findings also show significant differences compared to experimental data. Generally, simulations report lower melting temperatures than experimental results. For example, the reported experimental bulk melting temperatures for pure copper (Cu) and pure gold (Au) have been found to be 1358 K and 1338 K, respectively (Kittel, 1995). However, the simulation results obtained in this study were below these values. This discrepancy is due to the nature of the atomistic potentials used, the simulation parameters, and especially the nanoscale system size. The high surface-to-volume ratio of nanostructures causes surface energies to dominate, which decreases the melting temperature (Qi, 2004, Buffat, 1976). Consequently, the composition ratio of Cu-Au Janus nanoparticles exerts a significant influence on the resulting melting temperature. A substantial decrease in melting temperatures has been observed in systems where the Cu ratio decreases, concomitant with an increase in the Au content. This indicates that the thermal stability of the Janus structure is influenced not only by the type of metal but also by the ratios of the metals utilized.

### 3.2. Heat Capacity

Another critical parameter in analyzing thermal properties is heat capacity. Heat capacity is defined as the quantity that expresses the amount of heat required to increase the temperature of a substance by one unit (1 °C). It is employed as a significant parameter in atomistic simulations, particularly in the context of determining phase transitions and understanding the thermal stability of the system. The calculation of heat capacity at constant volume ($C_v$) is achieved through the following expression (Wang, 2007, Rapaport, 2004):

$$\frac{C_v}{k_B} = \frac{1}{Nk_B^2 T^2} (\langle PE^2 \rangle - \langle PE \rangle^2) + \frac{3}{2} \qquad (2)$$

In the given equation (2), the constant $k_B$ is the Boltzmann constant, $N$ is the total number of atoms in the system, $T$ is the given temperature, and $PE$ is the potential energy. Given the stability of the system volume, a constant contribution of $3/2\ k_B$ from kinetic energy is observed at each temperature.

In this study, the constant volume heat capacities ($C_v$) of Cu-Au JNPs have been investigated by molecular dynamics simulations for different composition ratios. The obtained $C_v$ data have proven to be instrumental in elucidating the thermal stability and phase transition behavior of nanoalloy systems. Increases and peaks in $C_v$ values that occur abruptly generally signify the melting point.

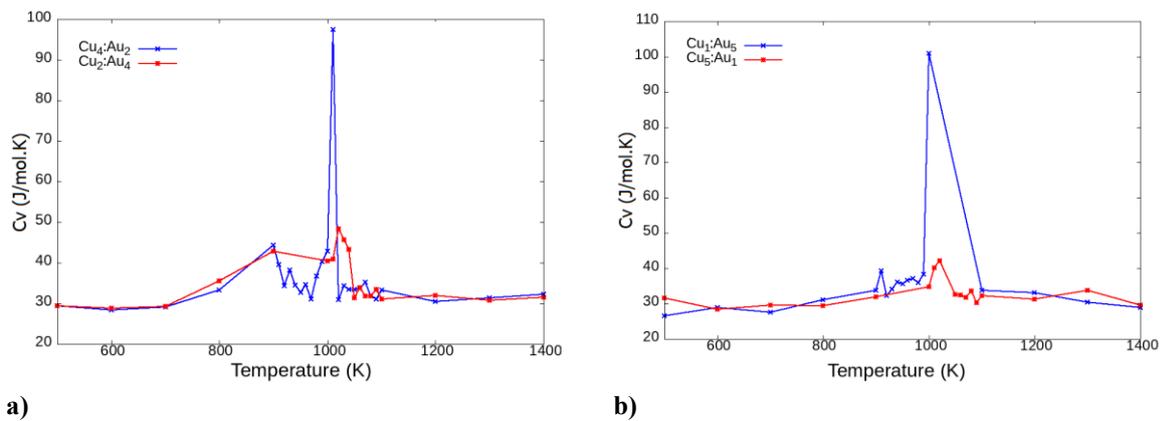

**Figure 4.** The temperature-dependent heat capacity of JNP's a. $Cu_4:Au_2$ and $Cu_2:Au_4$, and b. $Cu_1:Au_5$ and $Cu_5:Au_1$.

As illustrated in Figure 4a, the Cu4:Au2 model system with elevated Cu content demonstrates a significant maximum in its constant volume heat capacity (Cv) at approximately 1000 K. This observation indicates that the system exhibits a structural phase transition to a liquid state at this particular temperature. This observation indicates that the system undergoes a structural phase transition due to melting at this temperature. In a similar approach, the Cu2:Au4 composition has been observed to demonstrate a phase transition behaviour that is concomitant with the Cv curve around 1010 K. However, the peak value in this model system remains at lower levels. The substantial transition observed in the constant volume heat capacity ($C_v$) curve of the Cu2:Au4 model system depicted in Figure 4a indicates a decline in the system's thermal stability, resulting in a phase transition occurring over a wide temperature range. The phenomenon under investigation has been attributed to the high proportion of surface atoms and compositional irregularities in bimetallic nanoparticles (Qi, 2004). As shown in Figure 4b, the heat capacity of the $Cu_1:Au_5$ system has been observed to exhibit a significant increase around 1000 K, followed by a sharp decline. This finding suggests that the system demonstrates an earlier and more abrupt phase transition. Conversely, a broader transition range and a less pronounced $C_v$ peak are observed in the $Cu_5:Au_1$ system. This observation suggests that the system exhibits enhanced stability and more precise phase transition regulation.

These findings agree with the extant literature emphasising the decisive role of composition ratio on thermal stability in Cu-Au nanoalloys. As demonstrated by Qi and Wang (2004), an increase in the copper (Cu) content of Cu-Au alloys results in an elevated melting temperature of the system, consequently leading to the formation of more stable structures. This result is attributed to the higher bond energy and lower atomic mobility of Cu. In a review study, Aletto and Ferrando (2005) posited that the structural stability of nanoalloys is closely related to parameters such as the coordination number of atoms, surface energy and intrinsic symmetry. In conclusion, the obtained $C_v$ data demonstrate that the thermal stability of Cu-Au Janus nanostructures varies significantly depending on the composition; with an increase in Cu content, the nanosystems become more stable and show clear phase transitions.

### 3.3. Radial Distribution Function

The radial distribution function, denoted *g(r)*, is a statistical descriptor of the probability that another atom will be found at a distance *r* from a specified reference atom within atomistic systems. RDF facilitates the quantitative analysis of short- and medium-range atomic order, with a particular emphasis on amorphous, semi-crystalline, and nanostructured systems. RDF profiles, which are observed as prominent peaks in crystalline structures, are considered structural indicators reflecting degradation, melting transition, or reorganization of the structure with temperature (Frenkel, 2002).

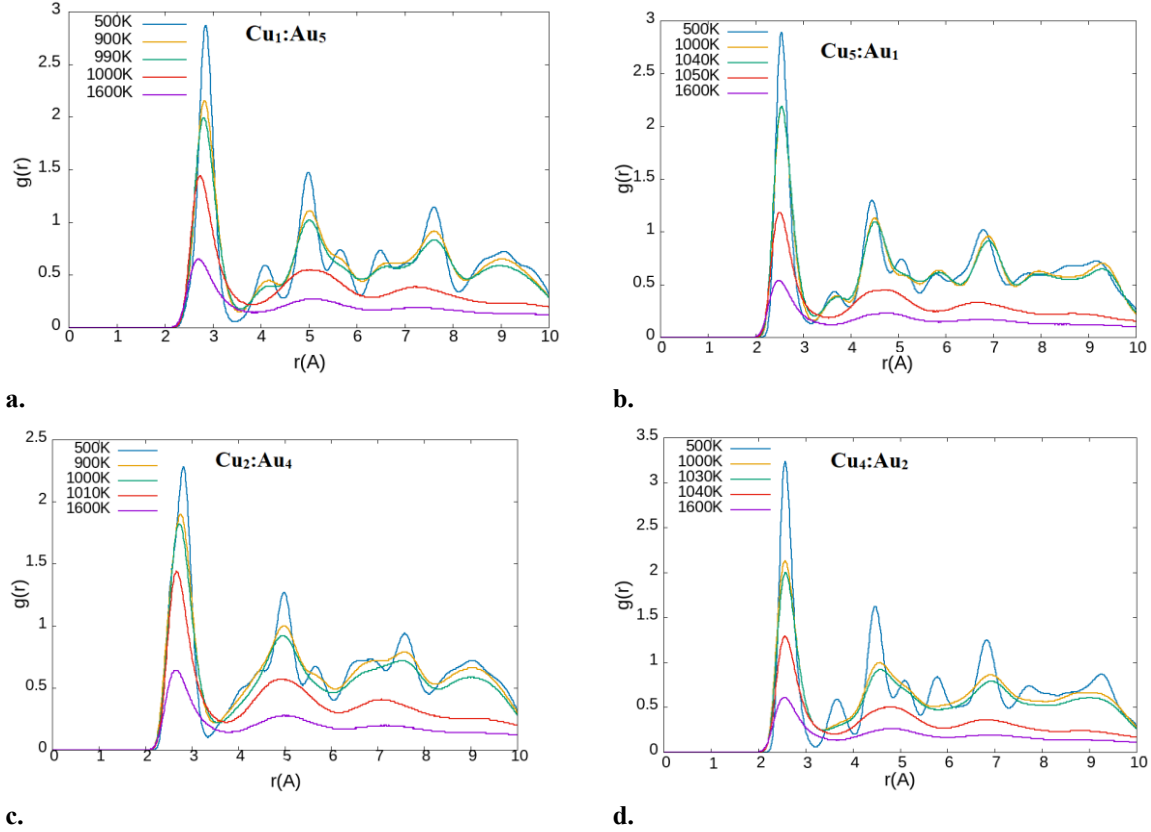

**Figure 5.** Radial distrubition functions of a) $Cu_1:Au_5$ b) $Cu_5:Au_1$ c) $Cu_2:Au_4$ and d) $Cu_2:Au_4$ Janus NPs with size of 5 nm, respectively.

The RDF for a system consisting of a single atom type is expressed as follows [64];

$$g(r) = \frac{V}{N} \frac{n(r)}{4\pi r^2 \Delta r} \quad . \tag{3}$$

In Equation (3), $N$ represents the total number of atoms in the system and $V$ is the system's volume. The function *n(r)* gives the number of atoms within a spherical shell of thickness *Δr*, while *4πr²Δr* represents the volume of this shell. The RDFs of the Cu-Au based NPs are computed to predict their melting temperature.

Figure 5a shows the RDF curves of a 5 nm diameter Janus NP with a Cu:Au ratio of 1:5 at temperatures of 500 K, 900 K, 990 K, 1000 K and 1600 K. This analysis has been carried out in order to understand both the crystalline order of the system and the temperature-dependent structural changes. At the low temperature of 500 K, the RDF function has distinct and sharp peaks. This finding suggests that the atoms are arranged in a highly ordered structure. The first coordination shell has been determined to be located at approximately r ≈ 2.7 A, which is consistent with the first neighbourhood distance of metals with FCC structure. As the temperature increases (particularly within the range of 900 K to 990 K), a substantial decrease in the amplitude of the peaks has been observed. The second and third coordination shells of the RDF also become increasingly faint, indicating that the crystal structure order of the model system has started to degenerate. The irregularities in the RDF of the model system increase at 1000 K. The peaks expand and their amplitude decreases, which indicates an increase in atomic vibrations and distortions within the structure. At 1600 K, the RDF has an irregular profile with no sharp peaks, indicating that the model system has largely melted. At this temperature, *g(r)* is approximately equal to 1, which indicates that the system has completely melted. The results obtained demonstrate that the crystal structure of $Cu_1:Au_5$ JNP becomes unstable with increasing temperature, and the melting transition occurs within the range of approximately 1000-1100 K. As demonstrated in Figure 5b, the initial and secondary coordination peaks for the $Cu_5:Au_1$ system exhibit notable stability up to temperatures ranging from 1040 to 1050 Kelvin. This observation indicates that the system maintains its structural integrity under conditions of elevated temperature. It is evident that, as the temperature rises above this threshold, two notable phenomena occur. Firstly, the peaks that were initially visible at greater distances begin to dissipate. Secondly, the initial peak undergoes an expansion in its dimensions. These observations are indicative of a phase transition within the structural composition. The melting temperature of this model system has been shown to be attributable to the predominance of Cu atoms, which exhibit a stronger bond structure and higher cohesion energy (Yin, 2008). As demonstrated in Figure 5c, the RDF peaks for $Cu_2:Au_4$ exhibit a significantly faster rate of decay with increasing temperature. At approximately 1000-1010 K, the initial peak demonstrates a substantial expansion. At 1600 K, the RDF profile is almost flat except for the first coordination shell, indicating that the structure has completely disordered to a liquid-like state. The earlier appearance of disorder in this composition can be explained by the lower cohesion energy of Au-rich systems (Qi, 2004). Figure 5d shows that the $Cu_4:Au_2$ nanoparticle exhibits intermediate behaviour. The RDF peaks persist until around 1030–1040 K, but their amplitude decreases gradually during this period. Compared to the $Cu_2:Au_4$ model system, this structure demonstrates greater thermal stability. This suggests that Cu atoms play a stabilising role in the structure of model systems (Wang, 2016).

In general, RDF analyses have shown that the thermal stability of model systems is directly dependent on their ratio. As the copper content increases, the nanoparticles retain their structural integrity at higher temperatures, and the disordering transition is delayed. This result reflects the impact of interatomic bonding strength, and is consistent with the findings of previous molecular dynamics studies of alloyed and Janus-type bimetallic nanoparticles (Li, 2013).

### 3.4. Lindemann Criterion

The Lindemann criterion is widely used to understand the thermal stability and melting processes of solids. It suggests that a crystal structure melts when the average vibration amplitude of its atoms exceeds a certain critical value. This criterion is an effective method for identifying nanoscale phase transitions, particularly in molecular dynamics (MD) simulations (Lindemann, 1910). The Lindemann parameter is defined by the following equation:

$$\delta_i = \frac{1}{N-1}\sum_{j \neq i} \frac{\sqrt{\langle R_{ij}^2 \rangle - \langle R_{ij} \rangle^2}}{\langle R_{ij} \rangle} \qquad (4)$$

In equation (4), $R_{ij}$ is the distance between atoms 'i' and 'j', and N is the total number of atoms in the model.

The parentheses indicate the average value taken across the system at the relevant temperature. The critical Lindemann value for melting is generally given as being in the range 0.1–0.2 in the literature (Jin, 2001, Belonoshko, 2006).

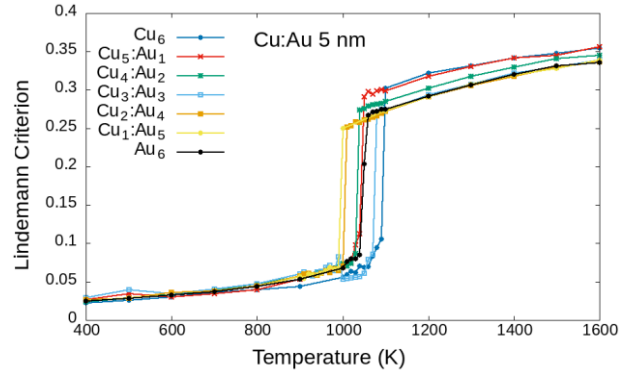

**Figure 6.** Lindemann Crtiterion of Cu:Au Janus NPs with sizes of 5 nm.

Figure 6 shows the temperature-dependent variations in the Lindemann parameter of pure $Cu_6$ and $Au_6$ model systems, as well as of JNPs with different Cu:Au ratios. A sudden increase is observed in all systems around 1000 K, indicating that the melting transition has occurred. This thermal approach is particularly favourable for $Cu_6$ and $Cu_5$: $Au_1$ systems and the Lindemann parameter reaches higher values for other JNPs. These results show that Cu atoms are more weakly bonded than Au atoms and therefore move more easily. It also explains why structural irregularities are observed earlier and at a lower melting temperature (Kofman, 1994, Alavi, 2006). It is observed that as the Cu content in the model systems increases, the melting temperature increases and the transition becomes sharper. In particular, the Lindemann parameter increases more steadily at lower temperatures in the $Cu_1$:$Au_5$ and $Au_6$ systems, resulting in a more abrupt transition. This is due to the higher bond energy and thermal stability of Cu atoms (Harischandra, 2020).

The differences in melting temperature and the structural changes in Janus-type NPs are related to their surface structure and Cu-Au ratio variation. The literature reports that increasing the amount of gold (Au) on the surface increases the melting temperature by decreasing the system's overall surface energy, whereas copper (Cu)-rich surfaces are more unstable (Belonoshko, 2006, Akbarzadeh, 2021). Compared to core-shell structures, it is said that Janus structures do not melt simultaneously in the surface and core regions, and compositional instabilities complicate the transition behaviour (Bochicchio, 2013).

Furthermore, analysis of the transition slopes in the Lindemann parameter curves reveals that an increase occurs in a wide temperature range before melting in model systems rich in Cu. In systems with a high number of Au atoms, however, this increase occurs in a narrower temperature range and more sharply. These results reveal that the melting mechanisms of these systems are strongly related to their composition (Akbarzadeh, 2021). These findings are in agreement with previous molecular dynamics studies showing that surface composition and structural irregularities have a decisive influence on the melting behaviour in Cu-Au bimetallic nanoparticles (Peng et al., 2015; Li et al., 2012).

### 3.5. Mean Square Displacement (MSD) and Diffusion Coefficient

MSD analysis is a fundamental tool for revealing nanoparticle temperature-dependent dynamics and phase transitions (Baletto, 2005). The MSD is the measure of how much the atoms have been displaced in relation to their initial positions in a given interval of time and is defined by the equation (5).

$$\langle r^2(t) \rangle = \langle (\vec{r_i}(t+t_0) - \vec{r_i}(t_0))^2 \rangle \qquad . \qquad (5)$$

In equation 5, $ri(t)$ represents the position of atom $i$ at time $t$. The linear increase that has been observed in the time-dependent MSD curves indicates that the system is diffusing (Frenkel, 2002). The diffusion constant D is calculated from Einstein's relation using this linear region (Allen, 1987):

$$D = \lim_{t \to \infty} \frac{\langle r^2(t) \rangle}{6t} \qquad . \qquad (6)$$

In this study, MSD calculations have been performed at different temperatures for Janus nanoparticles with different Cu:Au ratios of 5 nm diameter and the results are shown in Figure 7.

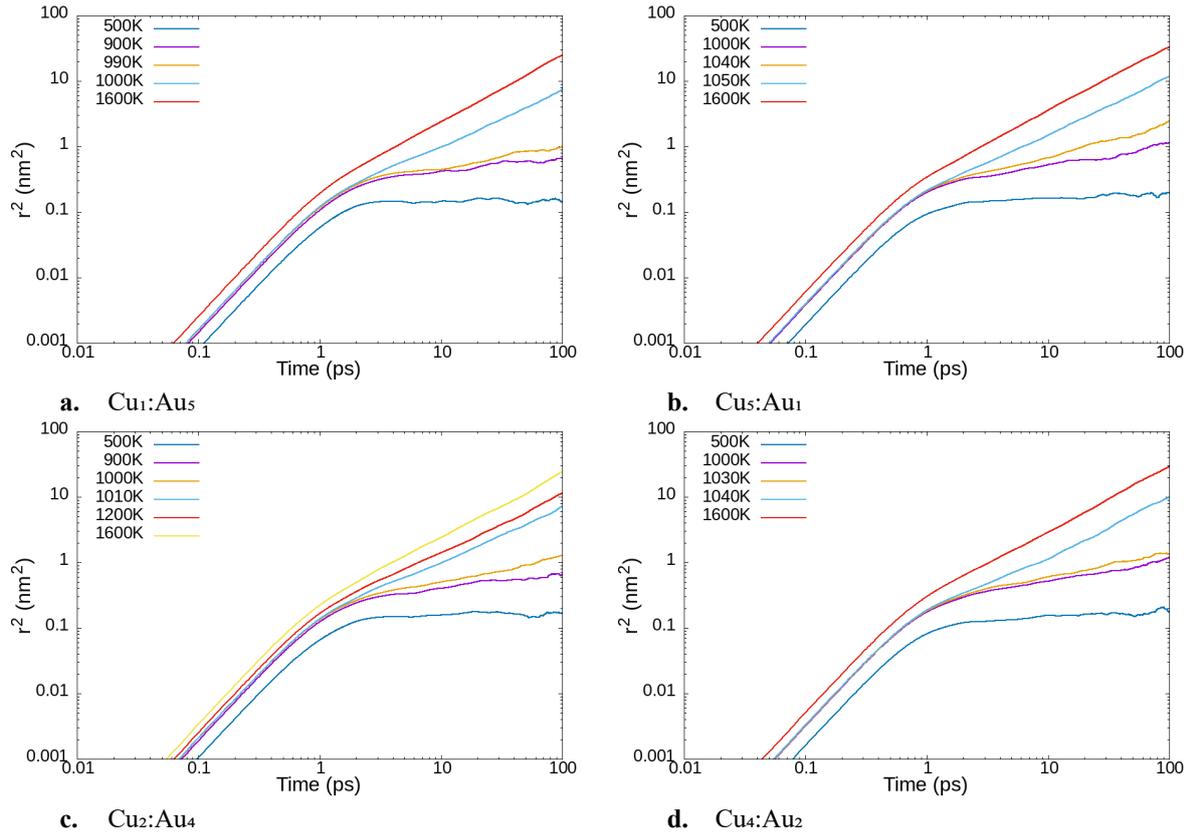

a. Cu$_1$:Au$_5$
b. Cu$_5$:Au$_1$
c. Cu$_2$:Au$_4$
d. Cu$_4$:Au$_2$

**Figure 7.** MSD curves obtained at different temperatures for JNP model systems.

Each graph shows that atomic displacement is restricted at low temperatures. This indicates that the system is in the solid phase and the atoms have mobility limited by lattice vibrations. The linearly increasing MSD curves with time, generally observed at temperatures above 1070 K and above, indicate that the atoms are randomly moving and the system exceeds the melting temperature. The convergence of the MSD curves obtained at 500 K, 900 K, 990 K and 1000 K for the Janus nanoparticle of Cu$_1$:Au$_5$ composition in Figure 7a to a constant value with time indicates that the atomic mobility is limited and the system retains its crystalline structure. This suggests that only small oscillations in the position of the atoms are realised below the melting temperature. The curve at 1000 K shows a small increase, suggesting that the system is close to a phase transition. In the Cu$_1$:Au$_5$ model system with rich Au component, it is known that Au atoms are enriched at the surface and contribute to the structural stability of the system (Li, 2016, Gafner, 2018). At the same time, the lower percentage of Cu atoms means that the system has limited mobility at low temperatures. Figure 7b shows the MSD curves of Cu$_5$:Au$_1$ JNP obtained at different temperatures. At low temperature (500 K, 1000 K) the MSD has limited increase with time, indicating that the system remains in the solid phase. At 1050 K the curves become clearly linear, indicating the beginning of atomic diffusion due to melting. The rapid and linear increase observed at 1600 K indicates that the system has completely transitioned to the liquid phase. These results suggest that the phase transition occurs around 1050 K.

Figure 7c shows that in Cu$_2$:Au$_4$ JNP, atomic motion remains limited in the range of 1000-1030 K, while at temperatures above 1040 K, diffusion behaviour becomes evident and melting begins. In Figure 7d, the melting temperature of the Cu$_2$:Au$_4$ model system occurs at 1040 K. It is already known from the literature that the melting temperatures of bimetallic nanoparticles can vary significantly, depending on the alloy composition and atomic arrangement (Qi, 2004). Low melting temperatures and surface-priority melting behaviour have been observed in gold-rich systems. This is related to the enrichment of gold atoms at the surface and the low surface energy (Lewis, 1997). In this context, the sudden increase in diffusion observed in the Cu$_2$:Au$_4$ system can be attributed to the earlier surface atom mobilization and collective melting of the structure. Furthermore, due to the high surface-to-volume ratio of the nanoparticles, they are observed to exhibit surface-assisted melting mechanisms rather than collective melting (Huang, 2013). These results suggest that the diffusion behaviour of Janus nanoparticles is strongly dependent not only on temperature but also on compositional asymmetry.

The temperature dependence of diffusion constants obtained from MSD analyses for model systems shows Arrhenius-type behaviour. The diffusion constant D is related to temperature by the Arrhenius equation:

$$D(T) = D_0 exp\left(-\frac{E_\alpha}{k_B T}\right) \quad (7)$$

Where $D_0$ is the self-diffusion coefficient, $E_\alpha$ is the activation energy, $k_B$ is the Boltzmann constant and $T$ is the absolute temperature. The $D_0$, $E_\alpha$ and $D$ values at 2000 K obtained for pure Cu and Au NPs and five different Cu-Au Janus NP compositions in this study are presented in Table 3.

**Table 3.** Arrhenius parameters of Cu-Au Janus model nanoparticle systems and diffusion constant $D$ values obtained at 2000 K.

| Model | $D_0$ (nm²/ns) | $E_\alpha$ (eV) | $D$ (nm²/ns) |
| --- | --- | --- | --- |
| $Cu_6$ | 40.12 | 0.264 | 8.64 |
| $Cu_5:Au_1$ | 18.87 | 0.246 | 4.52 |
| $Cu_4:Au_2$ | 52.25 | 0.326 | 7.90 |
| $Cu_3:Au_3$ | 36.58 | 0.295 | 6.60 |
| $Cu_2:Au_4$ | 35.08 | 0.296 | 6.30 |
| $Cu_1:Au_5$ | 29.13 | 0.270 | 6.07 |
| $Au_6$ | 24.81 | 0.248 | 5.86 |

The results in Table 3 show that the value of $D_0$ generally increases with increasing Cu content, but the activation energy $E_\alpha$ also increases. In particular, the $Cu_4:Au_2$ system has both the highest $D_0$ and the largest $E_\alpha$ value, revealing that a higher thermal activation is required for diffusion to start in this system. This finding indicates that, while Cu atoms facilitate enhanced atomic mobility, the system exhibits stronger atomic bonding (Lee, 2023). Conversely, within the $Cu_5:Au_1$ system, the activation energy is reduced, and diffusion commences at decreased temperatures. However, the diffusion constant value attained at 1000 K is lower than that of the other compositions. It has been demonstrated that with an increase in the Au ratio (towards the Cu1:Au5 system), there is a corresponding increase in the stability of the diffusion constants of the systems, and a decrease in the activation energies. This phenomenon can be attributed to the weaker binding energies of gold atoms, which have been shown to enhance their structural stability (Dong, 2011). These findings support the studies on nanoparticle melting mechanisms in the literature. As demonstrated in a number of studies, diffusion and melting events are initiated by surface atoms, particularly in small-sized systems. Furthermore, composition differences have been shown to significantly affect these processes (Kamachali, 2019, Lewis, 1997). In this context, the observation that Cu and Au atoms possess differing binding energies and surface activity gives rise to a composition-sensitive evolution of the temperature thresholds, diffusion characteristics and phase transition trends that are observed in Janus nanoparticles. It is evident from the obtained thermal behaviour profiles that not only temperature, but also atomic distribution and composition asymmetry, play a significant role in the melting process.

### 4. Results and Discussion

In this study, the thermal and structural properties of 5 nm Cu–Au Janus nanoparticles (JNPs) with varying compositions have been investigated through MD simulations using the embedded atom method (EAM). The main findings are summarized as follows:

- The melting process is characterized by a first-order phase transition from solid to liquid, identified by a sharp increase in total energy with temperature. Pure Cu nanoparticles exhibit the highest melting temperature (~1100 K), whereas pure Au nanoparticles melt at lower temperatures (~1060 K). Among Janus compositions, Cu-rich systems demonstrate higher melting points and enhanced thermal stability compared to Au-rich ones.

- Heat capacity ($C_v$) analysis reveals sharp peaks near the melting points of Cu-rich nanoparticles, indicating well-defined phase transitions. In contrast, Au-rich compositions show broader $C_v$ maxima, suggesting more gradual melting initiated at the surface.

- RDF results confirm that Cu-rich JNPs maintain crystalline order up to higher temperatures, while Au-rich particles lose structural order earlier due to weaker bonding.

- The Lindemann index increases sharply near the melting point, with Cu-rich systems displaying more abrupt solid-to-liquid transitions, reflecting stronger atomic cohesion.

- Diffusion behavior, analyzed via MSD and diffusion coefficients, follows Arrhenius temperature dependence. Cu-rich nanoparticles show higher activation energies and diffusion constants, indicating strong bonding coexisting with increased atomic mobility above melting.

Overall, these results demonstrate that the thermal stability and diffusion properties of Cu–Au Janus nanoparticles are strongly composition-dependent. Cu-rich compositions offer superior structural integrity and sharper phase transitions, whereas Au-rich systems are more thermally responsive but less structurally robust. These insights provide valuable guidance for designing bimetallic nanomaterials with tunable thermal properties for catalytic and electronic applications.